**ICMCTF 2013 conference**

# Predicting the steady state thickness of passive films in order to prevent degradations of implants.

Jean Geringer*[,a,b], [a], Matthew L. Taylor[b] , Digby D. Macdonald[b]

[a] Bio-Tribocorrosion lab, Biomechanics and Biomaterials Department

Center for Biomedical and Healthcare Engineering EMSE, CNRS:UMR5307, LGF,

158 cours Fauriel

F-42023 Saint-Etienne, France

Tel: (33) 477 426 688 / (1) 814 954 2225

E-mail: geringer@emse.fr

[b] Materials Science and Engineering

Center for Electrochemical Science and Technology

206A Steidle Building

University Park, PA 16802, USA

E-mail: jag54@psu.edu

**Abstract.**

Some implants have approximately a lifetime of 15 years. The femoral stem, for example, should be made of 316L/316LN stainless steel. Fretting corrosion, friction under small displacements, should occur during human gait, due to repeated loadings and un-loadings, between stainless steel and bone for instance. Some experimental investigations of fretting corrosion have been practiced. As well known, metallic alloys and especially stainless steels are covered with a passive film that prevents from the corrosion and degradation. This passive layer of few nanometers, at ambient temperature, is the key of our civilization according to some authors. This work is dedicated to predict the passive layer thicknesses of stainless steel under fretting corrosion with a specific emphasis on the role of proteins. The model is based on the Point Defect Model (micro scale) and an update of the model on the friction process (micro-macro scale). Genetic algorithm was used for finding solution of the problem. The major results are, as expected from experimental results, albumin prevents from degradation at the lowest concentration of chlorides; an incubation time is necessary for degrading the passive film; under fretting corrosion and high concentration of chlorides the passive behavior is annihilated.

## Introduction.

The number of hip prosthesis will increase in the future. It doubled from 2000 to 2010, in France [1]. In the USA, 250,000 hip prostheses annually are implanted in 2012. It is prospected that around 572,000 hip prostheses should be implanted in the USA in 2030. It is worth noting that the number of revisions of total hip arthroplasties (THA) should follow the same trend, i.e. doubling in 2030. Every year, one in every 30 Americans has hip prosthesis. Health issues on mobility of patients are related to people aging and younger and younger patients [2].

One of the big concerns related to lifetime of implants is of understanding mechanisms of wear for predicting stability and for avoiding debris production that is responsible of implants failure. When a prosthesis is implanted for the first time, the first step related to oseeointegration consists of adherence between bone tissues and the host material. After this step, due to stress shielding and the difference of mechanical properties between bone and metal alloys, debonding occurs and micro motions should appear between the femoral stem and the bone. This last factor should involve micro friction in liquid environment, called fretting corrosion [3]. As mentioned previously, even if metallic alloy has a bigger mechanical performance as the ones of bone, under fretting corrosion degradations the wear of metallic alloy is higher than the one of bone. Before investigating some tests close to the human case some fretting corrosion tests have been investigated between 316L SS samples, material related to hip implants, and PMMA (PolyMethylMethAcrylate), material close to bone by these mechanical properties [4,5]. Some interesting results have been provided by these experimental tests and the authors want to develop knowledge by adding a modeling approach adapted from the Point Defect Model [6-8]. The point consists of adding a part related to fretting, mechanical wear, to the predicting and deterministic PDM model. It is the first time, as experimental tests of fretting corrosion with mechanical and electrochemical monitoring processes, that this kind of modeling is suggested. Many experiments were investigated under fretting corrosion at open circuit potential. The following strategy was used for getting some information in order to apply the PDM added by fretting contribution.

Many hypotheses are necessary to combine the PDM related to the fretting corrosion case. The main one is the stationary conditions required for calculating the steady state thickness. Under fretting corrosion conditions the physical reality is not the steady state. It is assumed that the Electrochemical Impedance Spectroscopy diagram took time under 3 minutes, from $10^5$ Hz to 0.1 Hz on the spectral range, and the steady state conditions are supposed to be reached. The point is that the Open Circuit Potential, OCP, value did not change beyond 5 mV during the EIS acquisition i.e. the OCP value should be considered as the same.

This work aims at comparing the results coming from EIS experiments at different time and from the modeled ones for improving the parameters and the physical constants of the Point Defect Model. Thus finally, calculating the steady state thickness of the oxide layer is possible. It is worth noting that this comparison is possible because the anodic part is considered as the wear track area and the cathodic one is considered as the external part of the wear track area. This hypothesis is related to the fact that the majority of the current density is due to fretting contribution in the wear track area.

The expected steady state thickness will be considered as an indicator for knowing if the passive layer does exist or not with albumin, the thin solid film. From our opinion, knowing precisely the thickness of the oxide layer without any experimental measurement is not reasonable. However this modeling approach is a good insight for predicting the influence of experimental conditions on the passivity of materials.

## Materials and methods

The experimental aspect is well described in [4,5]. Thus some fretting corrosion tests were succeeded in order to study the effect of chlorides concentration cumulated with the concentration of albumin. The studied contact is 316L stainless steel, parallelepiped shape, against a cylinder of PMMA. The protocol of polishing has a huge importance and it is well described in [9]. The precise protocol is defined in [10]. The test solutions are composed with NaCl solutions with increasing concentrations of $10^{-3}$ mol.$L^{-1}$, $10^{-2}$ mol.$L^{-1}$, $10^{-1}$ mol.$L^{-1}$ and 1 mol.$L^{-1}$. 2 solutions will be considered in this study: without albumin and with 20 g.$L^{-1}$ in order to examine the effect and to model this one by the Point Defect Model. According to this experimental procedure one will pay attention on the elements required for modeling.

From experiments the expected results are the Nyquist diagrams coming from EIS measurements. During fretting corrosion tests, every 20 minutes, Nyquist diagrams, from EIS, were registered. The Figure 1 illustrated the registration process for one experiment at applied potential of -400mV/SCE and 20 g.$L^{-1}$ of albumin and 1M of NaCl. The Nyquist diagrams evolve during fretting corrosion test as highlighted in the Figure 1. The imaginary part, absolute value, increases with time that it should be due to the protective effect of albumin according to the time. From that point it should possible to compare experimental EIS graphs and modeled ones coming from adapted Point Defect Model.

In this section the modeling procedure will be described based on PDM. The Figure 2 summarizes the 7 reactions-PDM model [11]. This metal dissolution takes into account of the anodic part of the equivalent electrical circuit. The relation (1) describes the steady state thickness of the oxides layer on the metal surface. Only the reactions 3 and 7 are non conservative. Thus they are involved on the steady state thickness. The added W term is related to the mechanical term related to fretting. It is issued from experimental measurements and the term is defined by the Archard's law [12].

$$\frac{\mathrm{d}L}{\mathrm{d}t} = \Omega k_3^0\, \mathrm{e}^{a_3 V}\, \mathrm{e}^{b_3 L_{ss}}\, \mathrm{e}^{c_3 pH} - \Omega k_7^0\, \mathrm{e}^{a_7 V}\, \mathrm{e}^{c_7 pH} \left( \frac{C_{\mathrm{H}^+}}{C_{\mathrm{H}^+}^0} \right)^n - W = \frac{\mathrm{d}L^+}{\mathrm{d}t} - \frac{\mathrm{d}L^-}{\mathrm{d}t} - W \qquad [1]$$

The fundamental model is described on [10] and in [11] concerning the contribution of fretting on the stability of the passive layer.

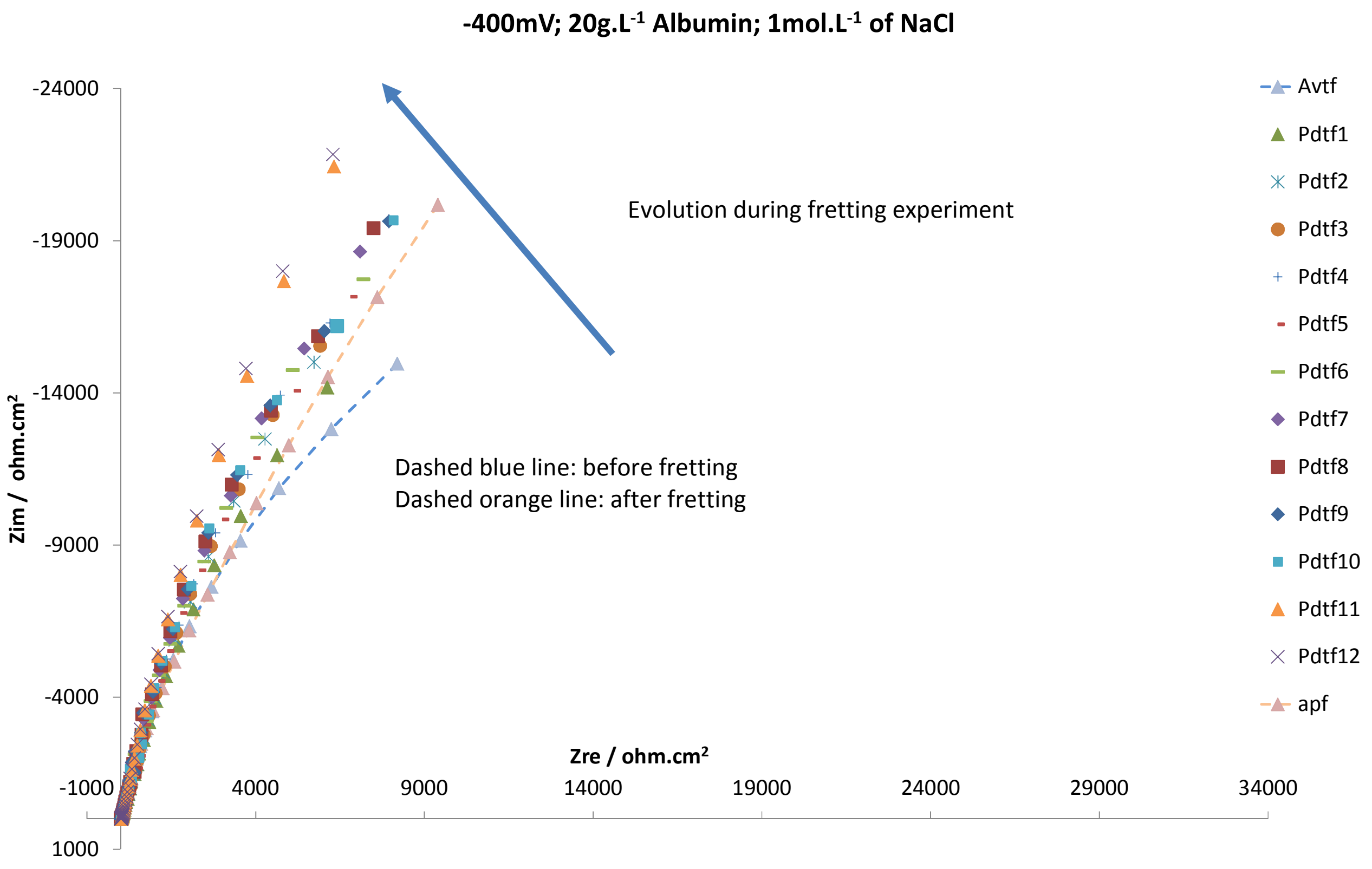

Figure 1: Nyquist diagrams related to an entire experiment of fretting corrosion, applied potential of -400mV/SCE, 1 mol.L$^{-1}$ of NaCl and 20 g.L$^{-1}$ of albumin. Avtf: before fretting test; Apf: after fretting test; Pdtf(i): during fretting corrosion experiment.

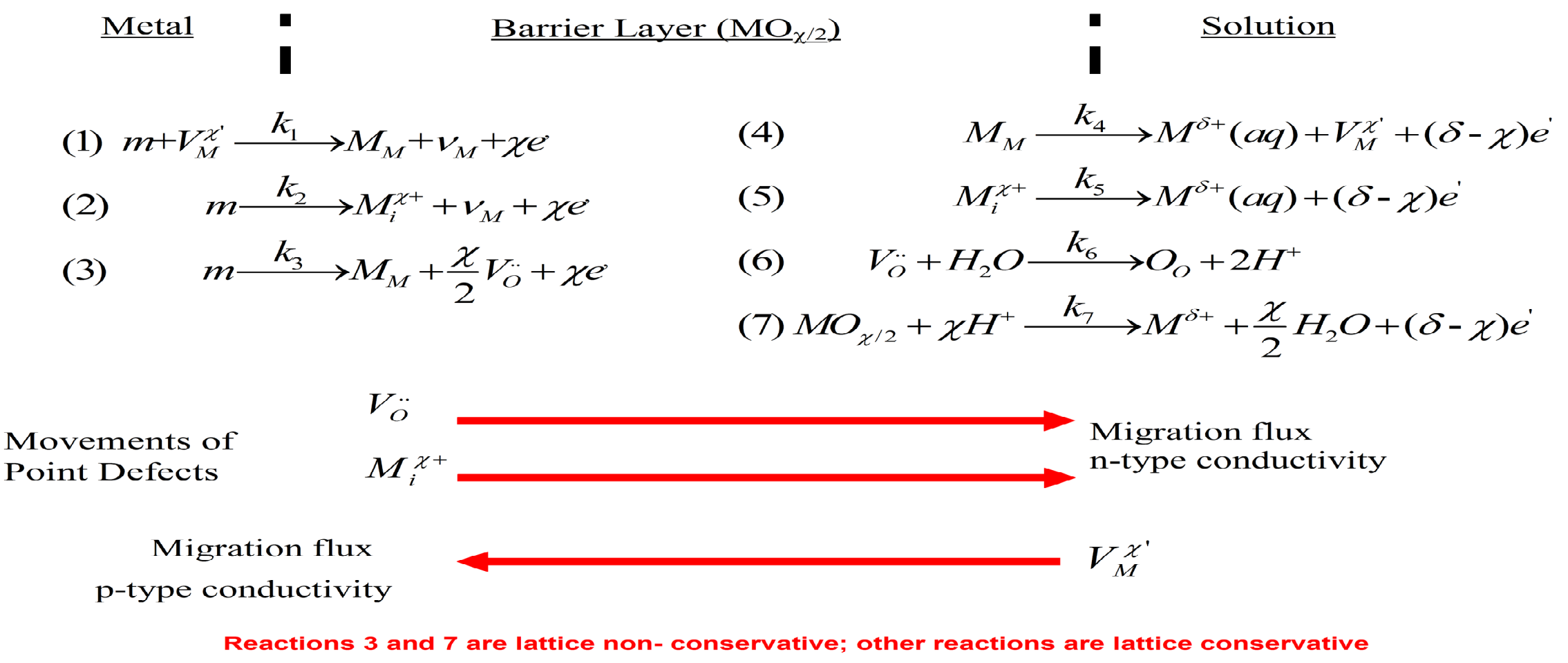

| | |
|---|---|
| $m$ : metal atom | $M^{\delta+}(aq)$ : metal cation in solution |
| $V_M^{\chi'}$ : cation vacancy on the barrier layer | $O_O$ : oxygen anion on the barrier layer |
| $M_M$ : metal cation on the barrier layer | $M_i^{\chi+}$ : interstitial cation |
| $V_M^{\chi}$ : metal vacancy | $V_{\ddot{O}}$ : oxygen vacancy on the barrier layer |

Figure 2: 7 reactions of the PDM, deterministic approach [7].

One coefficient related to Archard's law is adapted, fitted, according to the model. This approach describes the evolution of the passive layer on the 316L SS metal.

Thus for each EIS diagram, the potential is fixed then the current (or the current density) is measured. At that point, it can be calculated by the PDM (2):

$$I = F\left[\begin{array}{l} \chi k_1^0\, e^{a_1 V}\, e^{b_1 L_{ss}}\, e^{c_1 pH}\, C_{V_M}^{m/bl} + \chi k_2^0\, e^{a_2 V}\, e^{b_2 L_{ss}}\, e^{c_2 pH} + \chi k_3^0\, e^{a_3 V}\, e^{b_3 L_{ss}}\, e^{c_3 pH} + \\ (\delta - \chi) k_4^0\, e^{a_4 V}\, e^{c_4 pH} + (\delta - \chi) k_5^0\, e^{a_5 V}\, e^{c_5 pH}\, C_{M_i}^{bl/ol} + \\ (\delta - \chi) k_7^0\, e^{a_7 V}\, e^{c_7 pH} \left(\dfrac{C_{H^+}}{C_{H^+}^0}\right)^n \end{array}\right] \qquad [2]$$

The potential is imposed and the current density is measured, thus one may suggest that an equivalent electrical circuit could be suggested. This circuit consists of an anodic part (an impedance related to dissolution of metal, PDM, and a Warburg one), a cathodic part (parallel association of the oxygen reduction and the protons one) and a constant phase element related to the double layer on the surface of metal. These investigations have been succeeded till now. The innovation consists of using genetic algorithm for finding the 29 parameters coming from the Point Defect Model. The usual method, as the well-known Newton-Raphson, does not allow finding easily the global minimum. Advantages of the genetic algorithm are that best intermediate solutions are selected in the entire space of parameters. Finally the Figure 3 shows the experimental results of a Nyquist diagram and the fitted results provided by the model. The accordance sounds good, the relative discrepancy is less than 3%.

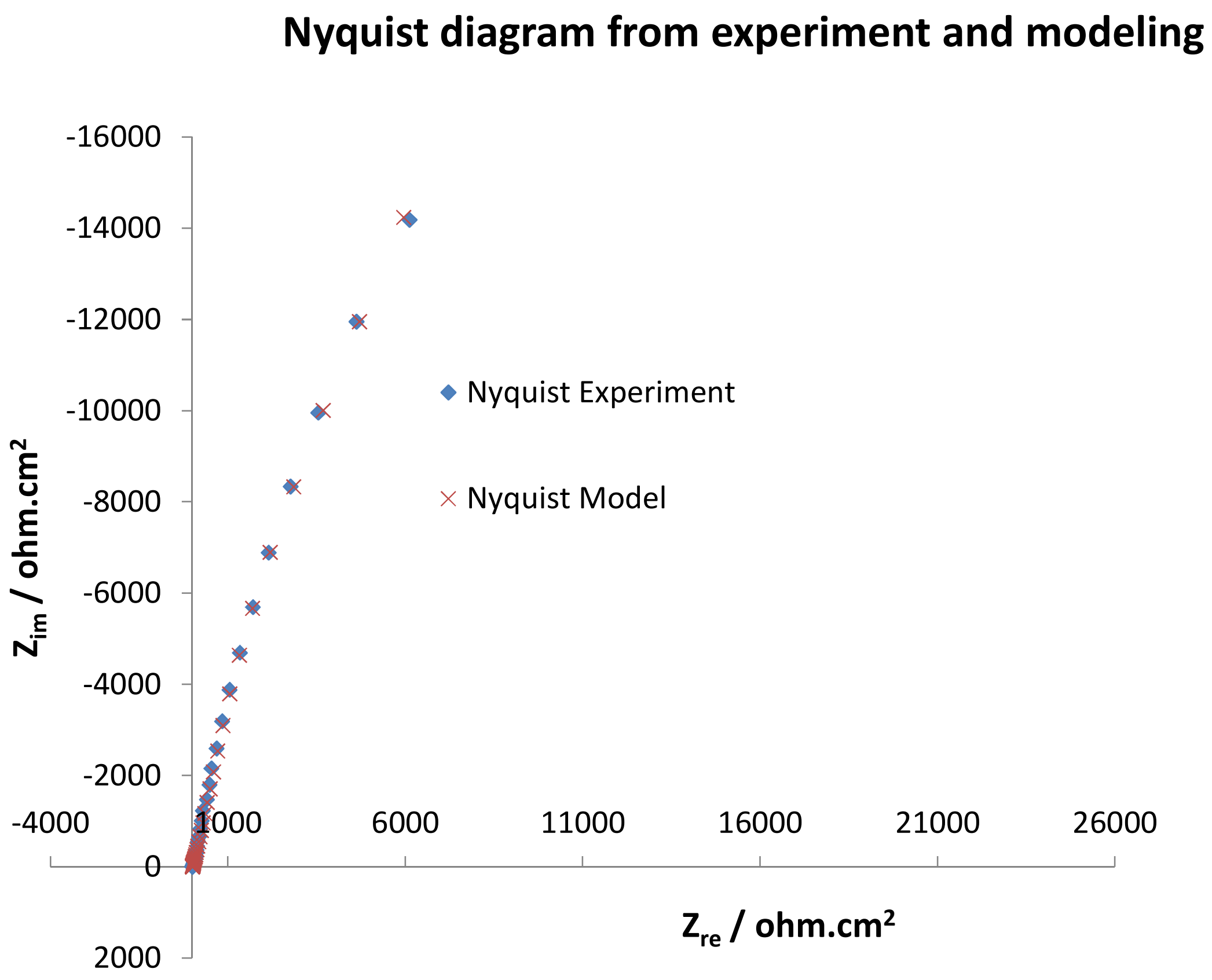

Figure 3: Nyquist diagrams related to one measurement coming from the experiment: 1 M of NaCl, 20 g.$L^{-1}$ of albumin and at applied potential of -400mV/SCE; the fitted points are modeling from the modeling based on the PDM with fretting corrosion contribution.

**Results and discussion.**

At Open Circuit Potential, OCP, the steady state thickness, Lss, should be calculated by the model. As mentioned the experiments related to Lss measurements have been not investigated. Thus the evolution of Lss has to be compared with experimental facts. The experimental evidence that could be compared with Lss evolution will be following the gaseous release from live video of the fretting contact area. Indeed this gaseous release should be reliable with an acidification of the contact area, as predicted in [10]. One will pay attention on Lss evolution according to the time during one experiment of fretting corrosion in fixed conditions. The Figure 4 shows two trends from Lss calculations issued from genetic algorithm procedure. Figure 4 a) [11] higlights the evolution of the modeled oxides thickness according to the time of one experiment. It is worth noting that an incubation time is necessary for observing the decrease of the oxides thickness. Beyond 100 minutes both phenomena appear simultaneously: from the modeling, as shown figure 4 a), the thickness is decreasing and from experiments the gaseous release is seen by video microscope registering. The experimental conditions are $10^{-2}$ $mol.L^{-1}$ of NaCl and 0 $g.L^{-1}$ of albumin. Concerning Figure 4 b), the NaCl concentration is the same than the one of Figure 4 a) but the concentration of albumin is 20 $g.L^{-1}$. On this figure, Lss is a little bit higher than the one without albumin. The incubation time seems to be the same. An experimental fact was that the gaseous release is less observed during the fretting corrosion experiment without albumin. Thus the modeled thicknesses and the experimental facts trend in the same way: albumin should protect 316L stainless steel from fretting corrosion degradations. This hypothesis confirms the evolution of the 316L SS wear volume: with proteins it is lower than the one without albumin [5].

a) b)

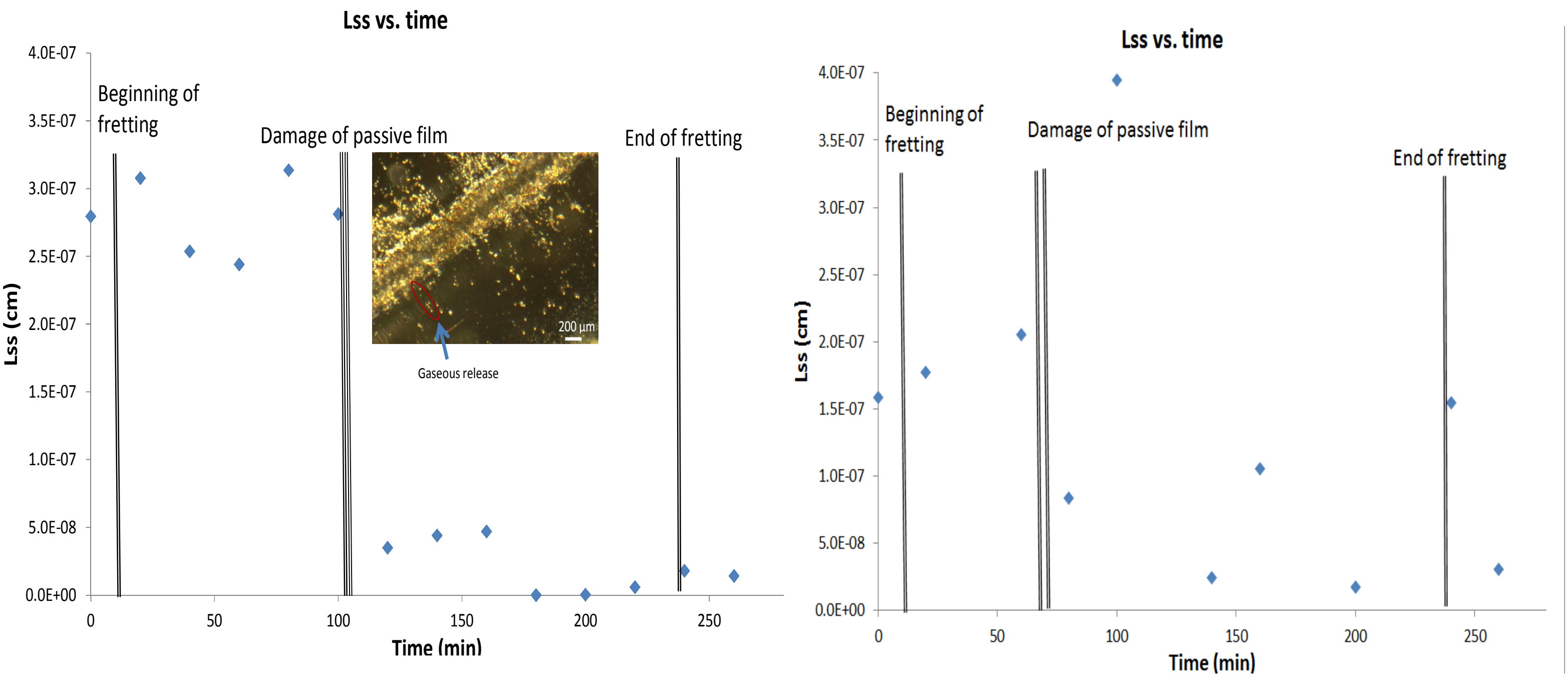

Figure 4: Oxides thickness of 316L SS according to the time during one fretting corrosion experiment, $10^{-2}$ mol.L$^{-1}$, 14,400 seconds of test; a) modeling without albumin, b) modeling with albumin.

The Figure 5 summarizes results related to the oxides layer thickness calculated thanks to the PDM. All the duration of the fretting corrosion is considered in this figure 5.

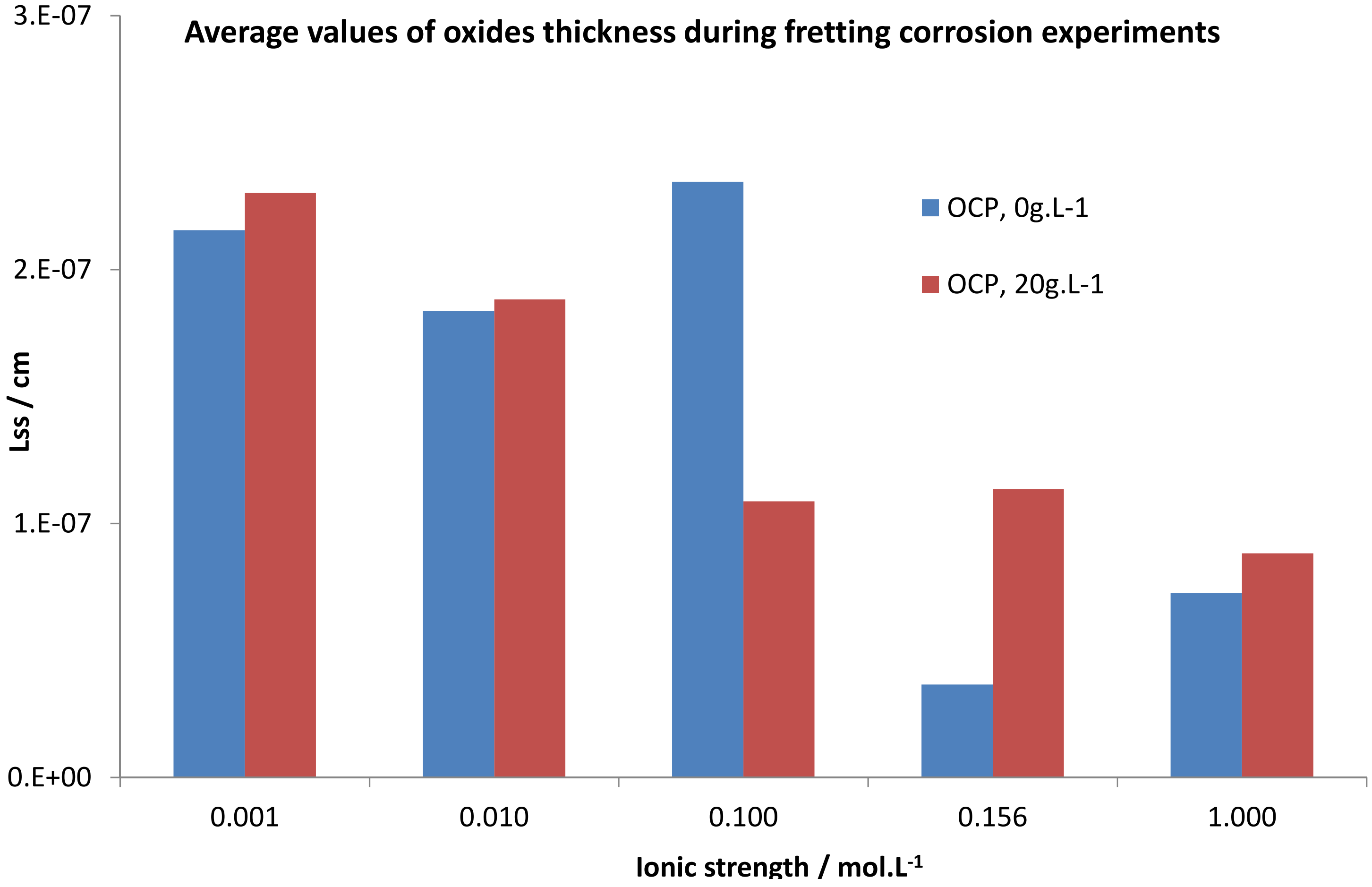

Figure 5: Average values of oxides layer calculated by the model based on PDM; 0.156 M of ionic strength is related to the Ringer solution (NaCl: 8.5 g.L$^{-1}$, KCl: 0.25 g.L$^{-1}$, $CaCl_2$,

$2H_2O$: 0.22 $g.L^{-1}$, $NaHCO_3$: 0.15 $g.L^{-1}$); these values are coming from 12 registered EIS diagrams, i.e. all time of fretting corrosion experiments.

The discrepancy, i.e. the confidence interval at 95%, of each experimental condition is so big. It is equal of 2 nm. As mentioned previously, the incubation time should be of 100 minutes. This factor should be taken into account for explaining the discrepancy that does not allow separating the influence of ionic strength and proteins. Thus it is the reason why the steady state thickness only is considered from 120 minutes of fretting corrosion experiment, Figure 6. Some tendencies can be extracted from results of Figure 6. First of all, at $10^{-3}$ $mol.L^{-1}$, albumin has a protective effect of the 316L SS passive layer. This result is in accordance with the lowest wear volume of 316L SS with 20 $g.L^{-1}$ of albumin, at this concentration of chlorides [5]. At $10^{-2}$ $mol.L^{-1}$, the calculated steady state thickness cannot separate the effect of albumin on the stability of the passive film. At 0.1 $mol.L^{-1}$ of ionic strength, i.e. 0.1 $mol.L^{-1}$of chloride ions in the solution, the trend seems reverse: albumin does not protect the metal. It is worth noting that this concentration is a threshold one and the model might be not efficient for predicting the steady state thickness. At 1 $mol.L^{-1}$, no influence of albumin can be extracted from data coming from modeling. This is the case of pure corrosion and the metal dissolution is active. The most interesting result concerns 0.156 mol.$L^{-1}$ of ionic strength, i.e. the Ringer solution. Obviously albumin is protecting the 316L SS metal. This solution is close by its constitution to the physiological liquid in the human body. One may conclude that albumin protects against fretting corrosion degradations of 316L SS.

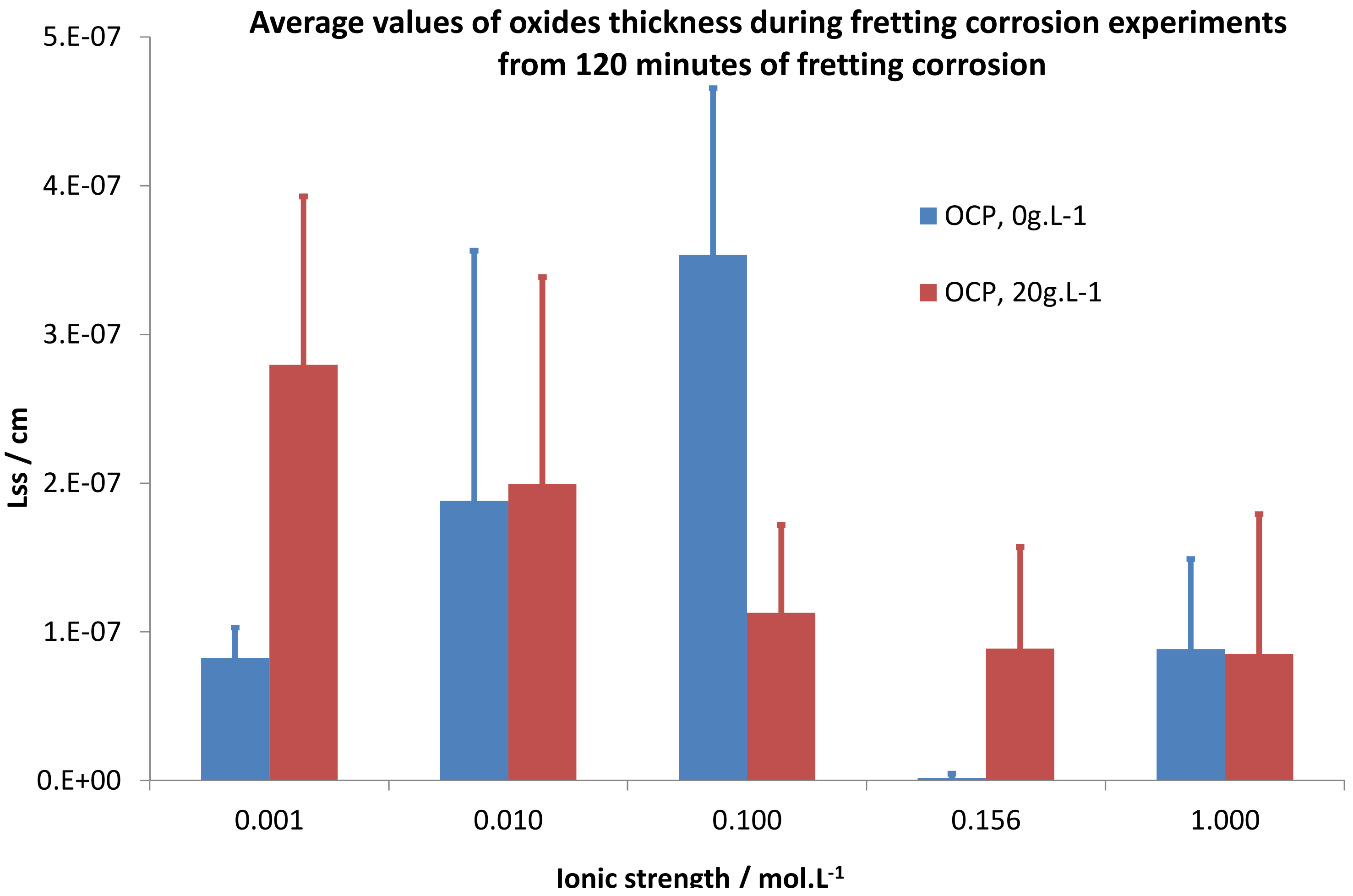

Figure 6: Average values of oxides layer calculated by the model based on PDM; 0.156 M of ionic strength is related to the Ringer solution (NaCl: 8.5 g.L$^{-1}$, KCl: 0.25 g.L$^{-1}$, $CaCl_2$, $2H_2O$: 0.22 g.L$^{-1}$, $NaHCO_3$: 0.15 g.L$^{-1}$); these values are coming from the last seven EIS measurements, i.e. after the incubation time of 100 minutes.

**Conclusions**

The steady state thickness has been calculated thanks to the Point Defect Model adapted to the case of fretting corrosion experiment. Two main points have been highlighted in this work. Firstly the incubation time of active dissolution can be estimated of 100 minutes of fretting corrosion test. The second point is that albumin protects 316L SS for the lowest concentration of chloride ions, i.e. $10^{-3}$ mol.L$^{-1}$, and in the case of Ringer solution. For 1 mol.L$^{-1}$ of chloride ions concentration, albumin has no effect on the steady state thickness of the passive layer of 316 SS. Finally for intermediate concentrations, i.e. $10^{-2}$ and $10^{-1}$ mol.L$^{-1}$, albumin cannot promote a higher protection of the passive film than the case of $10^{-3}$ mol.L$^{-1}$ of chloride ions in the medium. From these results coming from modeling, one might suggest that albumin protects 316L SS against fretting corrosion degradations for a concentration of $10^{-3}$ mol.$L^{-1}$. It is worth noting that the chloride ions concentration is between 0.15 and 1 mol.$L^{-1}$ in the physiological liquid, i.e. in the human body. Only taking into account the ions constitution of the physiological liquid, albumin has a beneficial effect on the 316L stainless steel degradation in the Ringer solution. Concerning the modeling process, thanks to the PDM, this is the first trial to adapt in the case of fretting corrosion. The genetic algorithm process is powerful for getting some interesting results on the influence of the specific coating that is albumin. Many prospects are under progresses: the modeling approach has to be improved, adapting the Point Defect Model to the specific case of fretting corrosion; the measurements of the oxides film thickness will be a point to investigate and the constitution. Fretting corrosion experiments are in progress with Co-Cr-Mo alloy and it will be a good candidate for applying the same methodology in order to estimate if the passive layer exists or not.

**Acknowledgements**

The authors would like to acknowledge the Rhône-Alpes Region / France, for grant which financed the 2011-2012 stay at Penn State University/PA/USA and Saint-Etienne Métropole / France, for financially supporting the experimental device and ENSM-SE for granting the Ph.D of Dr. Julie Pellier.

## References

1. HAS report, Evaluation des prothèses totales de hanche, Sept 2007.
2. S. Kurtz, K. Ong, E. Lau, F. Mowat, M. Halpern, the Journal of Bone and Joint Surgery Am, 89 (2007) 780
3. R.B Waterhouse, Fretting Corrosion, p. 4, Pergamon Press, Oxford (1975).
4. J. Geringer, B. Forest, P. Combrade, Wear 259 (2005) 943-951
5. J. Pellier, J. Geringer and B. Forest, Wear, 271 (2011) 1563
6. C.-Y. Chao, L.-F. Lin, D.D. Macdonald, J. Electrochem. Soc., 128 (1981) 1187
7. D.D Macdonald, Pure Appl. Chem., 71, 951 (1999)
8. D.D Macdonald, Electrochim. Acta., 56 (2011) 1761
9. J. Geringer, J. Pellier, M.L. Taylor, D.D. Macdonald, Thin Solid Films, 528 (2013) 123
10. J. Geringer, D.D. Macdonald, Electrochim Acta., 79 (2012) 17-30
11. J. Geringer, M.L. Taylor, D.D. Macdonald, ECS Trans., accepted in press.
12. J.F. Archard, J. Appl. Phys, 24 (1953) 981

**List of figures**

Figure 5: Average values of oxides layer calculated by the model based on PDM; 0.156 M of ionic strength is related to the Ringer solution (NaCl: 8.5 $g.L^{-1}$, KCl: 0.25 $g.L^{-1}$, $CaCl_2$, $2H_2O$: 0.22 $g.L^{-1}$, $NaHCO_3$: 0.15 $g.L^{-1}$); these values are coming from 12 registered EIS diagrams, i.e. all time of fretting corrosion experiments.

Figure 6: Average values of oxides layer calculated by the model based on PDM; 0.156 M of ionic strength is related to the Ringer solution (NaCl: 8.5 $g.L^{-1}$, KCl: 0.25 $g.L^{-1}$, $CaCl_2$, $2H_2O$: 0.22 $g.L^{-1}$, $NaHCO_3$: 0.15 $g.L^{-1}$); these values are coming from the last seven EIS measurements, i.e. after the incubation time of 100 minutes.